\documentclass[12pt]{article}
\pdfoutput=1
\usepackage{graphicx}
\usepackage{url}
\usepackage{rotating}
\usepackage{mathrsfs}
\usepackage{amssymb}
\usepackage{amsmath}
\usepackage{graphicx}
\usepackage{amsmath,amsfonts}
\usepackage{algorithm}
\usepackage{algorithmicx}
\usepackage{algpseudocode}
\usepackage{amscd}
\usepackage[mathscr]{eucal}
\usepackage{lineno}
\usepackage{extarrows}

\usepackage{listings}
\usepackage{tikz}

\numberwithin{thm}{section}
\newcommand{\be}{\begin{equation}} \newcommand{\ee}{\end{equation}}
\newcommand{\bd}{\begin{displaymath}} \newcommand{\ed}{\end{displaymath}}
\newcommand{\ba}{\begin{align}} \newcommand{\ea}{\end{align}}
\newcommand{\baa}{\begin{align*}} \newcommand{\eaa}{\end{align*}}
\newcommand{\ben}{\begin{enumerate}} \newcommand{\een}{\end{enumerate}}
\newcommand{\bi}{\begin{itemize}} \newcommand{\ei}{\end{itemize}}

\newcommand{\ud}{\mathrm{d}}

\newcommand{\var}[1]{\operatorname{Var}\left[ #1 \right]}

\newcommand{\cov}[2]{\operatorname{Cov}\left[ #1,#2 \right]}

\usepackage[normalem]{ulem}

\algnewcommand\And{\textbf{and}}

\begin{document}

%\linenumbers

\title{The phylogenetic effective sample size and jumps}
\author{Krzysztof Bartoszek} 

\maketitle

\begin{abstract}
The phylogenetic effective sample size is a parameter that has as its goal
the quantification of the amount of independent signal in a phylogenetically correlated sample.
It was studied for Brownian motion and Ornstein--Uhlenbeck models of trait evolution. Here,
we study this composite parameter when the trait is allowed to jump at speciation points
of the phylogeny. Our numerical study indicates that there is a non--trivial limit as
the effect of jumps grows. The limit depends on the value of the drift parameter of the
Ornstein--Uhlenbeck process.
\end{abstract}

%\subjclass[2010]{62B10; 62P10; 92--08; 92B10}

\section{Introduction: phylogenetic comparative methods with jumps}
Since its introduction to the evolutionary biology community \cite{JFel1988, THan1997}
the Ornstein--Uhlenbeck (OU) process 
\be
\label{eqSDEOU}
\ud X(t) = -\alpha(X(t)-\theta(t))\ud t + \sigma_{a}\ud B(t),
\ee
where $B(t)$ is the standard Wiener process, is the workhorse 
of continuous trait phylogenetic comparative methods. We can immediately notice that
taking $\alpha=0$ will result in the Brownian motion process 
(BM, popularized in the evolutionary biology community in \cite{JFel1985}).

These methods have as their aim 
the modelling of evolution of traits, like body size, on the between--species level. 
In particular this implies that the phylogenetic structure between the contemporary species
(providing the observations of the traits) has to be taken into account. 
The trait follows the stochastic differential equation model (e.g. Eq. \ref{eqSDEOU}) 
along each branch of the tree (with possibly branch specific parameters). 
At speciation times this process divides into two independently evolving processes. 

\begin{figure}[!ht]
\centering
\includegraphics[width=0.355\textwidth]{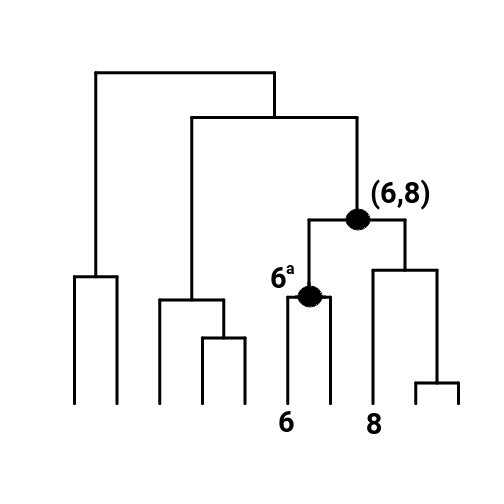}  
\includegraphics[width=0.355\textwidth]{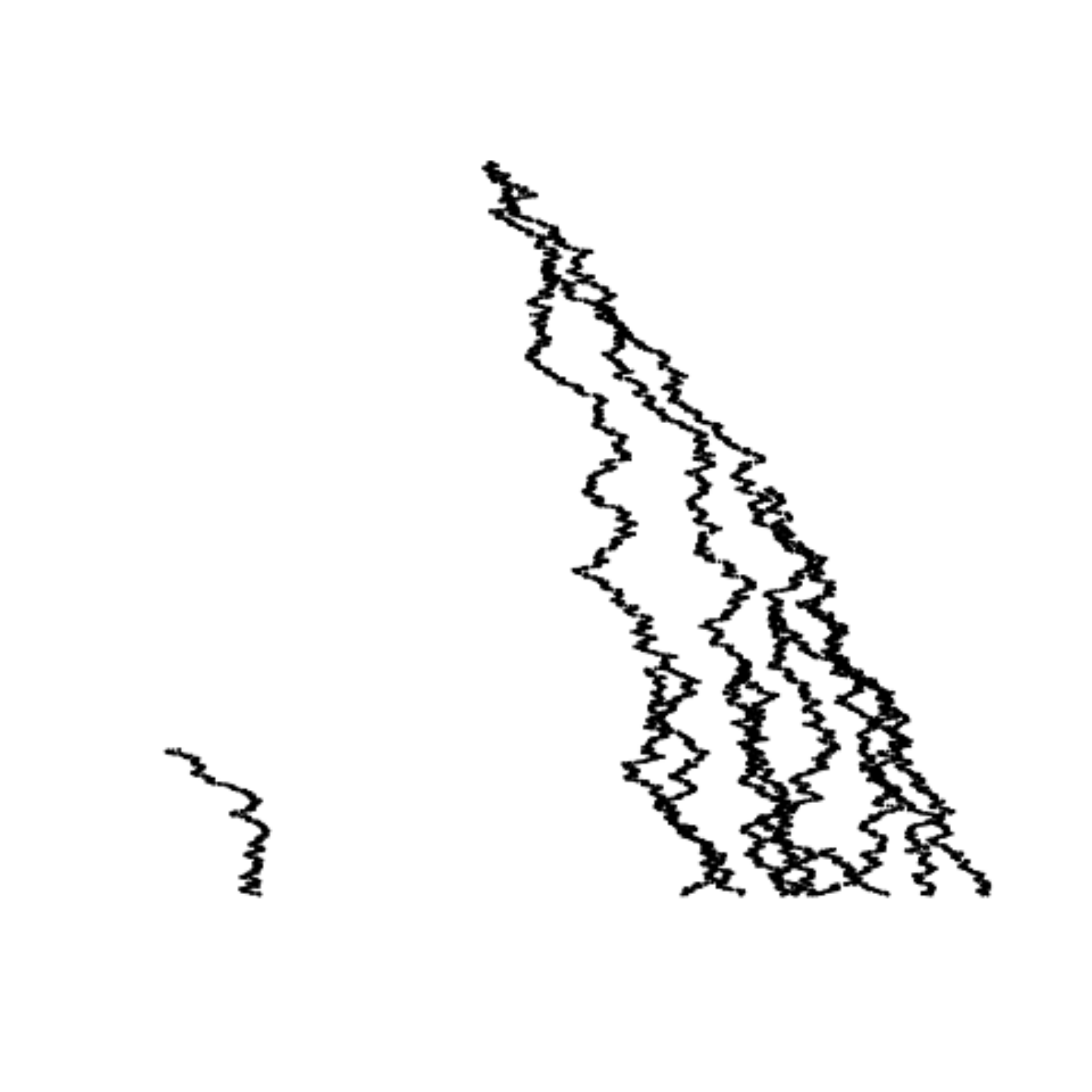} 
\caption{Left: tree with $10$ contemporary species. Right: OUj process evolving on this tree
(graphic by \texttt{mvSLOUCH} \cite{KBarJPiePMosSAndTHan2012}).
We can observe a single jump in the trait process.
% at the first speciation event after the root. 
The OUj process has parameters
%is a slowly adapting one with large jump variance 
$\alpha=0.3$, $\sigma_{a}^{2}=1$, $\theta=10$, $X_{0}=0$, $\sigma_{c}^{2}=8$, $p_{c}=0.05$ and the tree's height is $2.226$.
Time runs from top to bottom, i.e. the tree's tips are the contemporary species. 
}\label{figTree}
\end{figure}

In this work we will consider a variation of the OU model---the OU model with jumps (OUj). Just after speciation, 
independently on each daughter lineage, with constant (over the tree) probability $p_{c}$, a jump 
in the trait's trajectory can take place. The jump is normally distributed with $0$ mean and finite variance.
More formally if a speciation event takes place at time $t$, then, independently for each daughter lineage, the trait
value $X(t)$ will be 
\be\label{eqProcJ}
X^{+}(t) = (1-Z)X(t^{-}) + Z(X(t^{-})+Y).
\ee
By $X(t^{-/+})$ we mean the value of $X(t)$ respectively just before and 
after time $t$, $Z$ is a binary random variable with probability $p_{c}$ of being 
$1$ (the jump takes place ) and \mbox{
$Y\sim \mathcal{N}(0,\sigma_{c}^{2})$}, see Fig. \ref{figTree} for an example.

There is an evolutionary motivation for such a jump setup. If a species split into
two species, then this must have been the result of a dramatic event. 
A jump in an appropriate trait's value can catch rapid change associated with populations' division.
Of course, other jump models are possible. For example, only one daughter lineage can have a jump. This could 
correspond to a subpopulation breaking off from the main population. Here, we consider the simpler 
model of cladogenetic (at branching) evolution. 

Furthermore, the combination of jumps and an OU process seems to capture a key idea behind the theory of punctuated
equilibrium (i.e. the theory of evolution with jumps \cite{SGouNEld1977}). 
After a jump we could expect the trait to adapt very quickly, otherwise with a maladapted phenotype  
the species/population would be at a disadvantage. 
But as time passes, citing \cite{EMay1982} ``The further removed in time a species from the original
speciation event that originated it, the more its genotype will have become stabilized and the more
it is likely to resist change.''  Therefore, between the 
branching events (jumps) we could expect the trait to be stable---i.e. exhibit the phenomena
of stasis. However, stasis does not mean that the phenotype does not change at all,
rather that ``fluctuations of little or no accumulated consequence'' take place  \cite{SGouNEld1993}. 
This high level description is consistent with the OUj model. For $\alpha>0$ values the mean--centred, OU
process will converge with time to its stationary distribution---$\mathcal{N}(0,\sigma_{a}^{2}/(2\alpha))$.
Hence, what will be observed after a long time are stationary oscillations around the mean---and these
can be understood as statis between the jumps.

\section{The phylogenetic effective sample size}
From a statistical perspective the phylogenetically correlated trait sample 
is a collection of hierarchically dependent random variables. The dependency structure comes 
through the phylogeny. If the phylogeny was a star one, then all $n$ species would be independent
and if the phylogeny was degenerated to a star one but with all branch lengths equalling
$0$, all $n$ species would be identical and we would have $1$ observation. Between
these two extremes we have all phylogenies and one can ask whether one can 
measure the amount of independent observations in the observed sample. This
however, should not only depend on the phylogeny but also on the process driving the trait.
Ornstein--Uhlenbeck processes with large $\alpha$ lose information on the ancestral trait
faster and hence the contemporary observations should be ``more independent'' between
each other. %, than in the case of smaller $\alpha$s. 
A way to quantify this was suggested 
in (\cite{KBar2016}, following \cite{RSmi1994}), namely the \emph{phylogenetic effective sample size} (pESS) was defined
as $n_{e}= 1+p(n-1)$, where $p\in [0,1]$, with a number of proposals for $p$. We will focus
here on the \emph{regression effective sample} (rESS) approach for $p$. Let $\mathbf{V}$
be the $n\times n$ between--species--between--traits variance--covariance matrix, it 
depends both on the phylogeny and the process of evolution. We introduce the notation
that $\mathbf{V}_{i,}$ means the $i$--th row of $\mathbf{V}$ and 
$\mathbf{V}_{-i,}$ means $\mathbf{V}$ without the $i$--th row, analogously 
for columns $\mathbf{V}_{,i}$, $\mathbf{V}_{,-i}$. Define now $p$ as
\be \label{eqRegpESS}
p = n^{-1}\sum\limits_{i=1}^{n}\left(1-\mathbf{V}_{i,i}^{-1}\mathbf{V}_{i,-i}\mathbf{V}_{-i,-i}^{-1}\mathbf{V}_{-i,i} \right).
\ee

We can recognize this as the average scaled conditional variance, when regressing each species on all the others. 
Or in the language of linear regression, this is the average variance of the residuals, where each
residual comes from regressing each species on all the others. If we now have a normal model, like the OU,
then all such residuals will be independent and this can be used to quantify the amount
of independent signal in the phylogenetic sample. If all the observations are independent, $p=1$
and if identical $p=0$. 

At this point it should be emphasized that Eq. \eqref{eqRegpESS} is inspired by
the \emph{mean} effective sample size (mESS) considered in \cite{CAne2008}. There it is simply called 
the effective sample size and is defined as
\be n_{e}^{\mathrm{E}} = \vec{1}^{T}\mathbf{R}^{-1}\vec{1},\ee
where $\mathbf{R}$ is the correlation matrix derived from $\mathbf{V}$ and $\vec{1}$ is a
vector of $n$ ones. The parameter $n_{e}^{\mathrm{E}}$ is actually
the number of independent random variables that result in the same precision for
estimating the mean value (intercept) of a linear with $n$ correlated,
by $\mathbf{R}$, observations \cite{CAne2008}. It should be pointed out that the 
word ``mean'' in the name is only used to distinguish from the rESS
and is not connected to any average of sample sizes. 
%The usage of ``mean'' is motivated by the fact that $n_{e}^{\mathrm{E}}$ quantifies the information available on the mean value in a linear model.

\section{Simulation setup and results}
A key question in the study of punctuated equilibrium is whether it can be detected
based only on observing the contemporary sample. In \cite{FBok2002} it is noticed
(but only for the BM model) that such detection should be possible. 
The aim here is to study how the regression pESS reacts to 
the presence of jumps in the trait's trajectory. 

It is conjectured (based on a detailed numerical analysis and shown in BM case \cite{KBar2014}) 
that jumps in OU models of evolution cause a decrease of the interspecies correlation coefficient, $\rho_{n}$.
The interspecies correlation coefficient is defined \cite{SSagKBar2012} as the ratio of the
covariance between a randomly chosen pair of tip measurements with the variance of a randomly
chosen tip species. 
As the former covariance and variance are not conditional
on the tree, the $\rho_{n}$ parameter is a theoretical property of the tip values' distribution and is not sample
specific. However, treating it as a proxy for the amount of independent signal 
in the sample, its decrease would suggest an increase in the pESS.

In all simulations we fix $\sigma_{a}^{2}=1$, $X_{0}=\theta=0$ and $p_{c}=1$.
We consider all pairs $(\alpha,\sigma_{c}^{2})$, where
$\alpha \in  \{0,0.05,0.1,\ldots,0.95,1,1.25,1.5,1.75,2.25,2.5,5\}$
and
$\sigma_{c}^{2}\in \{0,0.1,\ldots,1.9,2,2.5,3,\ldots,9.5,10,11,\ldots,29,30,40,\ldots,90,100\}.$
The choice of the levels of $\alpha$ is motivated by the fact that the phylogeny
will be modelled by the pure--birth process with speciation rate $\lambda=1$. 
It is known (e.g. \cite{RAdaPMil2015,KBar2014,KBarSSag2015b}) that %for branching OU process
a qualitative behaviour phase transition takes place at $\lambda=\alpha/2$. Hence we 
explore Brownian motion $(\alpha=0)$, slow--adaptation $\alpha$--values $(\alpha<0.5)$, the critical value $(\alpha=0.5)$
and fast adaptation $\alpha$--values $(\alpha>0.5)$. The ancestral state $X_{0}$ and $\theta$ are not of interest to us
as they do not influence the variances and covariances. At the second moments level
$p_{c}$ and $\sigma_{c}^{2}$ appear only as their product (Appendix A.2 \cite{KBar2014})
so we are free to fix one of them to $1$, we choose $p_{c}=1$. The diffusion parameter, $\sigma_{a}^{2}$, is also
not of interest as it only enters through the ratio $\sigma_{a}^{2}/(2\alpha)$. 

We first simulate $10000$ Yule trees with $200$ contemporary species using the \texttt{TreeSim} \cite{TreeSim2} 
R  package. For each parameter set we independently subsample (due to lengthy running times) a collection
of $100$ trees. For each of these trees we calculate the between--tip--species covariance matrix.
The formulae for the variance of the trait value of tip species $i$, $X_{i}$,
under the OUj model can be recursively expressed  (cf. Appendix A.$2$ in \cite{KBar2014})
\be
\var{X_{i}} = \frac{\sigma_{a}^{2}}{2\alpha}(1-e^{-2\alpha t_{i}}) + e^{-2\alpha t_{i}}p_{c}\sigma_{c}^{2}
+e^{-2\alpha t_{i}}\var{X_{i}^{a-}},
\ee
where $X_{i}^{a-}$ is the value of the trait at the last branching event on the lineage to tip $i$
and $t_{i}$ is the length of the branch between this ancestor and tip $i$. The minus sign in the superscript
is to underline that $X_{i}^{a-}$ is the value at speciation and hence \emph{before} the jump took place.
In Fig. \ref{figTree} if we took tip species $i=6$, then $X_{i}^{a-}$ would be the value at the internal node
labelled $6^{\mathrm{a}}$ and $t_{i}$ the length of the branch between tip $6$ and node $6^{\mathrm{a}}$.
On the other hand, as jumps take place after speciation, independently on each daughter lineage, the 
covariance between the trait measurements, $(X_{i},X_{j})$, of the pair of tip species  $(i,j)$ will simply be
\be
\cov{X_{i}}{X_{j}} = e^{-2\alpha(T-\tau_{i,j})}\var{X_{i,j}},
\ee
where $X_{i,j}$ is the value of the trait at the node corresponding to the most recent common ancestor
of tips $i$ and $j$, $T$ is the height of the tree and $\tau_{i,j}$ is the time that passed from this 
most recent common ancestor to today. In Fig. \ref{figTree} if we took 
the pair of tip species $i=6$ and $j=8$, then $X_{i,j}$ would be the value at the internal node
labelled $(6,8)$ and $\tau_{i,j}$ the sum of branch lengths on the path between tip $6$ (equivalently tip $8$)
and node $(6,8)$. Because a jump happens just \textit{after} speciation, any jumps associated with 
this ancestral node are not shared by the pair of tip species and hence cannot contribute
to the covariance between them.

\begin{figure}[!ht]
\centering
\includegraphics[width=0.355\textwidth]{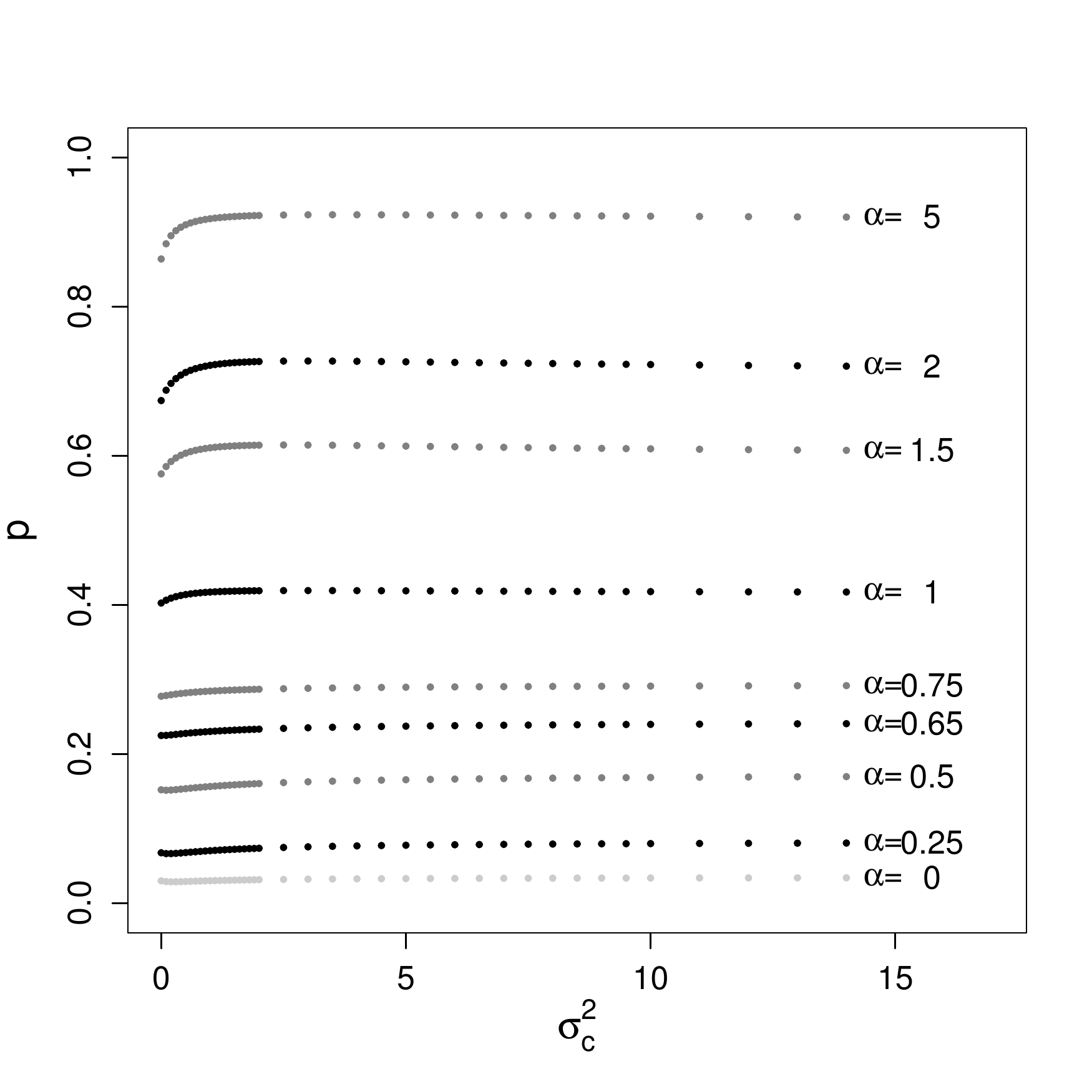} %CZY \includegraphics[width=0.45\textwidth]{regpESSfactorYuleSamp.pdf
\includegraphics[width=0.355\textwidth]{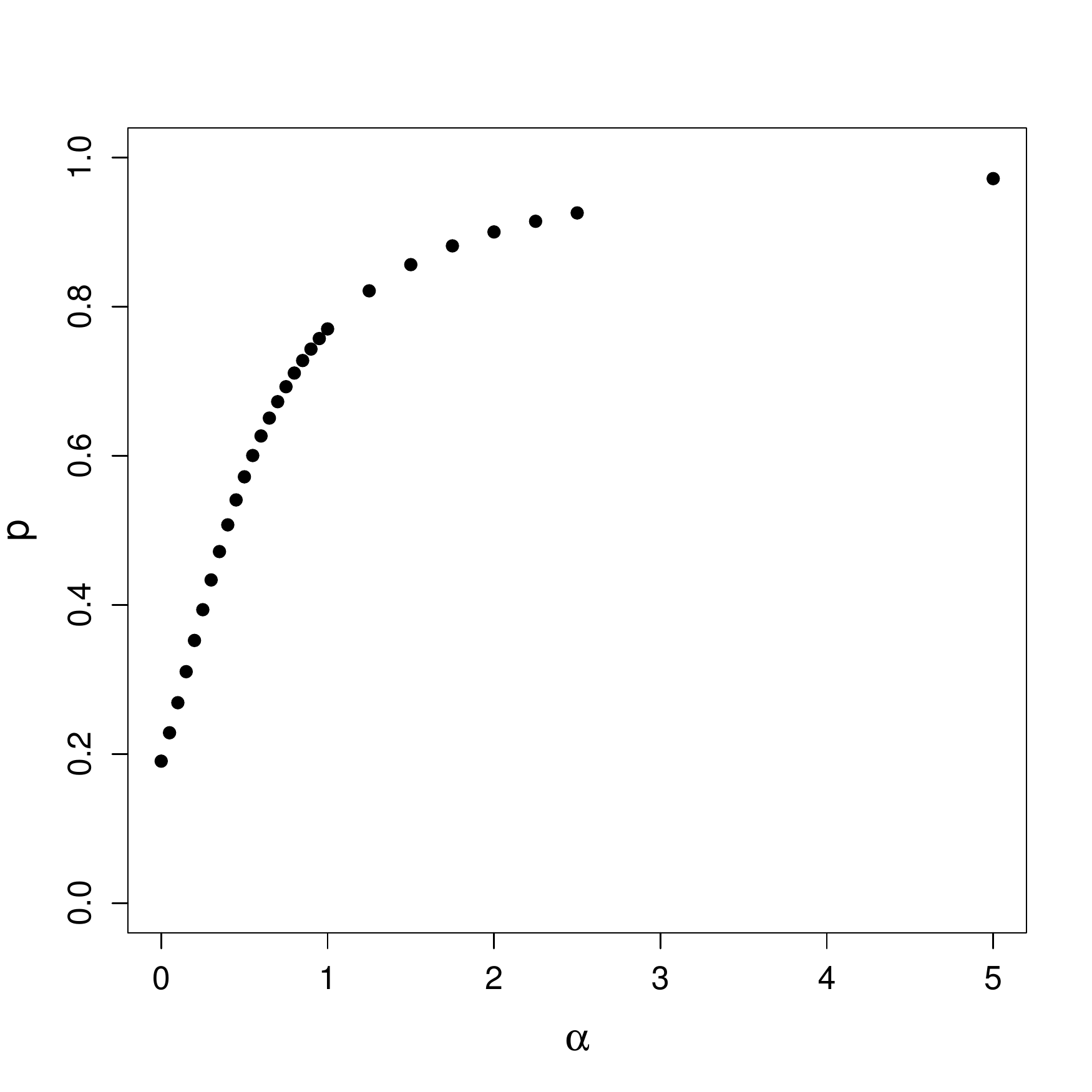} %CZY \includegraphics[width=0.45\textwidth]{regpESSfactorYuleSamp.pdf
\caption{Average 
(from $100$ trees for each point) values of the mean (right) and 
regression (left) pESS factors for the different model setups. The trees were all Yule trees with speciation rate equalling $\lambda=1$.
The values are presented for a representative subset of all the $\alpha$ values. 
All simulations were done in R version 3.4.2 \cite{R} running on an openSUSE 42.3 (x86\_64) box.
%, hence $\alpha=0.5$ is a critical value where a phase transition in the model's behaviour takes place. For $\alpha<0.5$ the OU process is slowly adapting (or tree is branching quickly) while $\alpha>0.5$ is the fast adaptation (or slow branching) regime. 
\label{figRespESS}
}
\end{figure}

From each calculated covariance matrix (notice that we do not need to simulate any trait trajectories) 
we calculate the mean and regression phylogenetic effective sample sizes 
and then report the average (over the $100$ Yule trees) effective sample size factor, $p$, for the given parameter
set. In Fig. \ref{figRespESS} we plot the observed results, factors derived from rESS and from mESS. 

\begin{figure}[!ht]
\centering
\includegraphics[width=0.355\textwidth]{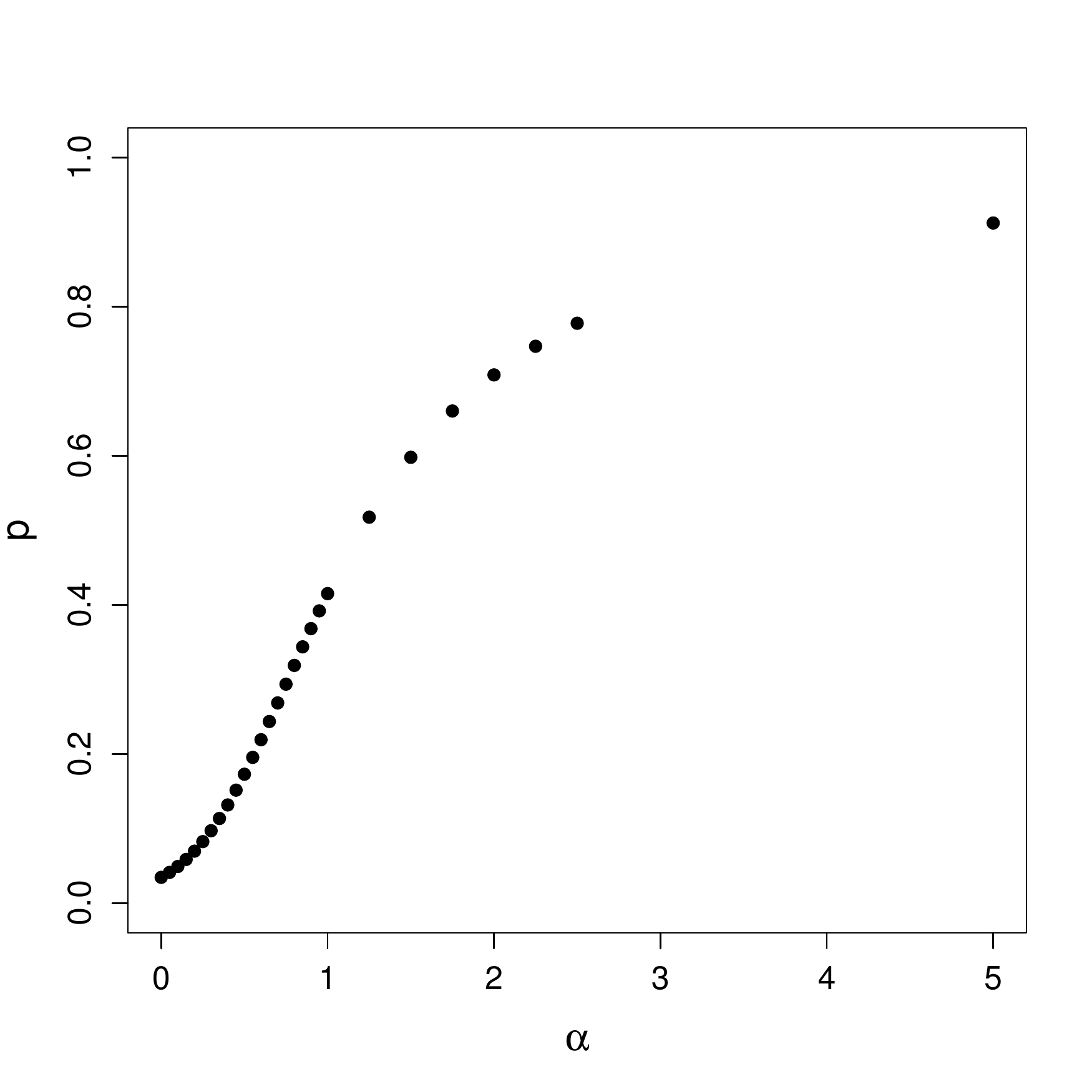} %CZY \includegraphics[width=0.45\textwidth]{regpESSfactorYuleSamp.pdf
\includegraphics[width=0.355\textwidth]{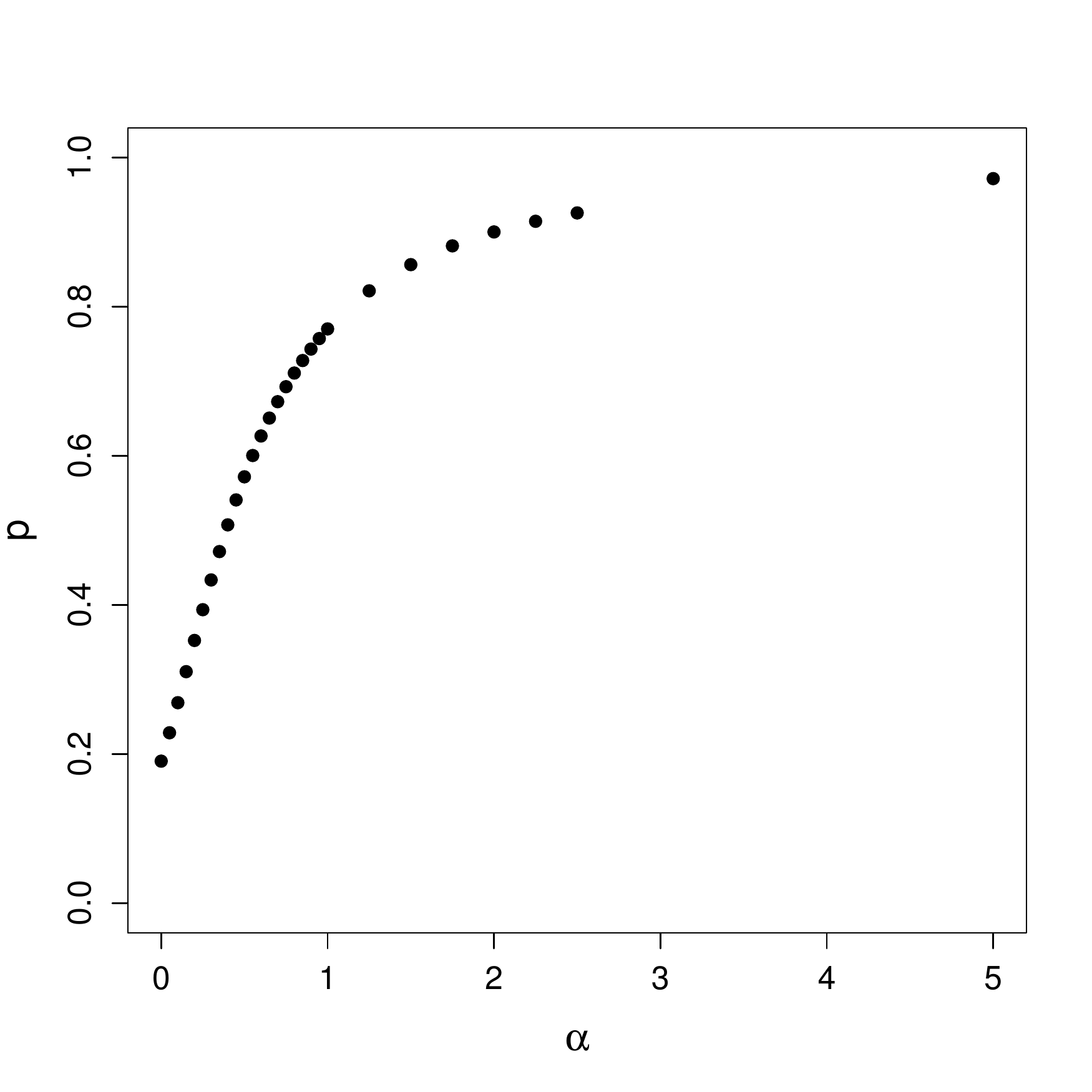} %CZY \includegraphics[width=0.45\textwidth]{regpESSfactorYuleSamp.pdf
\caption{Average 
(from $100$ trees for each point) values of the mean (right) and 
regression (left) pESS factors for different values of $\alpha$ when $\sigma_{c}^{2}=100$.
%The trees were all Yule trees with speciation rate equalling $\lambda=1$.
%The values are presented for a representative subset of all the $\alpha$ values. 
%All simulations were done in R version 3.4.2 \cite{R} running on an openSUSE 42.3 (x86\_64) box.
%, hence $\alpha=0.5$ is a critical value where a phase transition in the model's behaviour takes place. For $\alpha<0.5$ the OU process is slowly adapting (or tree is branching quickly) while $\alpha>0.5$ is the fast adaptation (or slow branching) regime. 
\label{figRespESSLimit}
}
\end{figure}

The results of the numerical study, plotted in Fig. \ref{figRespESS}, are in agreement with intuition.
The analysis of the theoretical model parameter, the interspecies correlation coefficient $\rho_{n}$,
indicates simple dynamics---as the influence of jumps increases, $\rho_{n}$ decreases 
(see Figs. 4 and 6 of \cite{KBar2014}). This suggests more independence in the sample
and hence a greater value of $p$, as observed in Fig. \ref{figRespESS}. 
As the value of $\sigma_{c}^{2}$, increases so do the pESSs. 
It is not surprising that $p$ increases with $\alpha$. The larger $\alpha$ is the quicker
the OU process looses information on the ancestral state. %Hence, $p\to 1$ with $\alpha \to \infty$.

The interesting conclusion from Fig. \ref{figRespESS} is that the rESS factors seem to have
a non--trivial limit, i.e. neither $0$ nor $1$, as $\sigma_{c}^{2}$ increases. Non--trivial limits with $n$ (without jumps)
were observed in \cite{KBar2016}, see Fig. 2 therein. Furthermore, convergence to this limit seems rapid. 
Interestingly, there does seem to be any visible dependency of this speed of convergence on $\alpha$.
At the first moment level there is a phase transition in the
Central Limit Theorems for the average of the contemporary sample \cite{RAdaPMil2015,KBarSSag2015b,KBar2016arXiv}
at $\alpha=\lambda/2$. In the mESS factors' case rapid convergence seems present when $\alpha \ge 1$,
but for lower values of $\alpha$ the situation is not obvious. The limits (or rather case when $\sigma_{c}^{2}=100$)
for different levels of $\alpha$ are plotted in Fig. \ref{figRespESSLimit}. For the rESS the points
seem to fall on a concave curve. In contrast, in the mESS's situation the underlying curve seems
to have an inflection point near $\alpha=0.5$, supporting that the phase 
transition influences the mESS. %However, nothing like this is evident for the rESS.

The mESS factors are lower than the rESS ones. This is again consistent with \cite{KBar2016} and intuition.
The mESS captures only information content on the mean, while the rESS all independent signal. The
observed data contains information not only on the mean but also on the second moments. When $\alpha$ is 
small the mESS factor is much smaller. This again agrees with intuition from the non--jump case,
as for small (relative to $\lambda$) $\alpha$
``local correlations will dominate over the ergodic properties of the Ornstein--Uhlenbeck process''
\cite{RAdaPMil2015}, to the extent that one cannot consistently estimate the root
state in the BM case, i.e. $\alpha \to 0$ limit \cite{CAne2008,KBarSSag2015a,SSagKBar2012}.

An applied motivation for undertaking this study is whether the pESS has any hope of being 
able to detect if the jumps have a large effect on the trait's evolution. The rESS factors
presented in Fig. \ref{figRespESS} do give hope for this. If one can point on which curve level
the rESS should lie, then one would estimate the parameters of the OU process (without jumps,
parameter identifiability with jumps present is not clear yet). If the estimated
parameters point to the same curve, then one can hypothesize that jumps do not play much of a role.
However, if they point to a curve above the expected one, then this indicates jumps. 
One can argue how one can know on which curve the rESS should lie for a given clade.
In principle this cannot be known, but it could be inferred (with some degree of error) from
sister clades, provided that the biological assumption of similar speeds of adaptation holds.

The numerical study presented here points to interesting mathematical directions of work. 
Firstly, how to characterize the limits of the pESS factors as the jumps dominate and then to find the speed
of convergence to these limits. Finally, the proposed in the previous paragraph method to detect punctuated 
equilibrium deserves its own detailed study.

\section*{Acknowledgments}
KB's was supported by Vetenskapsr\aa dets grant no. $2017$--$04951$ and
the Stiftelsen f\"or Vetenskaplig Forskning och Utbildning i Matematik.


\begin{thebibliography}{9999}
\bibitem{RAdaPMil2015} R. Adamczak, P. Mi\l o\'s, \textit{{CLT} for {O}rnstein--{U}hlenbeck branching particle system}, Elect. J. Probab. 20, 1-35 (2015). 

\bibitem{CAne2008} C. An\'e, \textit{Analysis of comparative data with hierarchical autocorrelation}, Ann. Appl. Stat. 2, 1078-1102 (2008).

\bibitem{KBarJPiePMosSAndTHan2012} K. Bartoszek, J. Pienaar, P. Mostad, S. Andersson, T. F. Hansen, \textit{A phylogenetic comparative method for studying multivariate adaptation},
J. Theor. Biol. 314, 204-215 (2012). 

\bibitem{KBar2014} K. Bartoszek, \textit{Quantifying the effects of anagenetic and cladogenetic evolution},
Math. Biosci. 254, 42-57 (2014). 

\bibitem{KBarSSag2015a} K. Bartoszek, S. Sagitov,  \textit{A consistent estimator of the evolutionary rate.},
J. Theor. Biol. 371, 69-78 (2015). 

\bibitem{KBarSSag2015b} K. Bartoszek, S. Sagitov,  \textit{Phylogenetic confidence intervals for the optimal trait value},
J. App. Probab. 52, 1115-1132 (2015).  

\bibitem{KBar2016} K. Bartoszek, \textit{Phylogenetic effective sample size},
J. Theor. Biol. 407, 371-385 (2016). 

\bibitem{KBar2016arXiv} K. Bartoszek, \textit{A Central Limit Theorem for punctuated equilibrium},
ArXiv e-prints 1602.05189 (2016). 

\bibitem{FBok2002} F. Bokma, \textit{Detection of punctuated equilibrium from molecular phylogenies}, J. Evol. Biol. 15, 1048-1056 (2002).

\bibitem{JFel1985} J. Felsenstein, \textit{Phylogenies and the comparative method}, Am. Nat. 125, 1-15 (1985). 

\bibitem{JFel1988} J. Felsenstein, \textit{Phylogenies and Quantitative Characters}, Annu. Rev. Ecol. Syst. 19, 445-471 (1988). 

\bibitem{SGouNEld1977} S. J. Gould, N. Eldredge, \textit{Punctuated equilibria: the tempo and mode of evolution reconsidered}, Paleobiology 3, 115-151 (1977). 

\bibitem{SGouNEld1993} S. J. Gould, N. Eldredge, \textit{Punctuated equilibrium comes of age}, Nature 366, 223-227 (1993). 

\bibitem{THan1997} T. F. Hansen, \textit{Stabilizing selection and the comparative analysis of adaptation}, Evolution 51, 1341-1351 (1997). 

\bibitem{EMay1982} E. Mayr, \textit{Speciation and macroevolution}, Evolution 36, 1119-1132 (1982). 

\bibitem{R} R Core Team, {\it R: A Language and Environment for Statistical Computing},
R Foundation for Statistical Computing, Vienna; www.R-project.org, 2013.

\bibitem{RSmi1994} R. J. Smith, \textit{Degrees of freedom in interspecific allometry: {A}n adjustment for the effects of phylogenetic constraint},
Am. J. Phys. Anthropol. 93, 95-107, (1994). 

\bibitem{TreeSim2} T. Stadler, \textit{Simulating Trees with a Fixed Number of Extant Species},
Syst. Biol. 60, 676-684 (2011).

\bibitem{SSagKBar2012}  S. Sagitov, K. Bartoszek, \textit{Interspecies correlation for neutrally evolving traits},
J. Theor. Biol. 309, 11-19 (2012).

\end{thebibliography}
\end{document}